\documentclass[twocolumn,showpacs,prb]{revtex4}
\usepackage{amssymb}
\usepackage{makeidx}
\usepackage{amsmath}
\usepackage{graphicx}
\usepackage{dcolumn}
\usepackage{bm}
\usepackage{epstopdf}

\begin{document}

\title{Impact of electron-electron Coulomb interaction on the high harmonic
generation process in graphene}
\author{H.K. Avetissian}
\author{G.F. Mkrtchian}
\email{mkrtchian@ysu.am}
\affiliation{Centre of Strong Fields Physics, Yerevan State University, 1 A. Manukian,
Yerevan 0025, Armenia}

\begin{abstract}
Generation of high harmonics in a monolayer graphene initiated by strong
coherent radiation field, taking into account electron-electron Coulomb
interaction is investigated. A microscopic theory describing the nonlinear
optical response of graphene is developed. The Coulomb interaction of
electrons is treated in the scope of dynamic Hartree-Fock approximation. The
closed set of integrodifferential equations for the single-particle density
matrix of a graphene quantum structure is solved numerically. The obtained
solutions show the significance of many-body Coulomb interaction on the high
harmonic generation process in graphene.
\end{abstract}

\pacs{78.67.Wj, 72.20.Ht, 73.22.Lp, 42.50.Hz}
\date{\today }
\maketitle

\section{ Introduction}

In the last decade, graphene and its analogs have attracted enormous
interest due to their unique electronic and optical properties.\cite{Castro}
The potential of graphene as an effective nonlinear optical material has
triggered many nonlinear optical studies. In particular, graphene-like
nanostructures can serve as an active medium for nanolasers and frequency
multipliers.\cite{RM} In their original structure freestanding graphene is
centrosymmetric and even-order nonlinear effects\cite%
{2nd1,2nd2,2nd3,2nd4,2nd5} vanish within the dipole approximation. The
latter is fully justified for perpendicular incidence of a pump wave to the
graphene plane, and the symmetry-allowed odd-order nonlinear optical effects
are very strong in graphene. This is confirmed by the experimental\cite%
{3rde1,3rde2,3rde3} and theoretical\cite%
{3rd1,3rd2,3rd3,3rd4,3rd5,3rd6,3rd7,3rd8} investigations of the third
harmonic generation process. With the increase of the pump wave intensity
one can enter into the regime\cite{Mer1,Mer2,Mer} where multiphoton effects
are essential and high-harmonics are generated, which until last decade have
been the prerogative of atomic systems.\cite{HHG} There are several
theoretical investigations devoted to the high harmonics generation in
graphene\cite{HH1,HH2,HH3,Mer1,Mer2,Mer3,HH4,HH5,HH6,Mer4,Mer5,HH7,HH8}
which have not yet validated experimentally.\cite{HH1-exp,HH2-exp,HH3-exp}
Note that the last experiment\cite{HH3-exp} with generation of ninth
harmonic in graphene opens new avenue towards the high-harmonic generation
in nanostructures, meanwhile earlier experiments reported about weak signals
at harmonics\cite{HH2-exp} or even absence\cite{HH1-exp} of nonlinear
response in the THz range of frequencies.

Various approaches have been made in the investigation of high harmonics
generation in graphene, however, all these investigations rely on
noninteracting electron picture and these results are applicable when the
electron-electron Coulomb interaction effects are negligible. In this
context, it is worthy to note that in 2D nanostructures it is possible to
tune the electron-electron Coulomb interaction by choosing the dielectric
constant of the substrate on which 2D material is deposited. Thus, depending
on the substrate material one may achieve a situation when electron-electron
interaction effects are essential and can significantly modify graphene
physics.\cite{FV1,FV2,FV3,E1,E2,E3,E4,E5,E6} As was shown experimentally\cite%
{FV1,FV2} and theoretically,\cite{FV3} in contrast to the single-particle
picture, the real spectrum of graphene is nonlinear near the neutrality
point, and Fermi velocity describing its slope, increases significantly.
Even an excitonic condensate along with the opening of an energy gap has
been predicted.\cite{E1,E2,E3,E4} Electron-electron interaction also
significantly modifies linear optical response of graphene.\cite%
{clop1,clop2,clop3,koch,K1,K2,K3} Hence, it is of interest to clear up the
influence of electron-electron interaction on the nonlinear optical response
of graphene, which is the subject of the present investigation.

In the present work, we develop a nonlinear microscopic theory of a
monolayer graphene interaction with the coherent electromagnetic radiation
taking into account the electron-electron Coulomb interaction using the
self-consistent Hartree-Fock approximation\cite{koch2} that leads to a
closed set of integrodifferential equations for the single-particle density
matrix. We neglect the scattering processes which are described by the
second-order terms in the carrier-carrier interaction\cite{K3}.\textrm{\ }%
Thus, we consider nonlinear coherent interaction in the ultrafast excitation
regime when relaxation processes are not relevant.\textrm{\ }Since we are
interested in both inter and intraband transitions for the light-matter
interaction Hamiltonian, we use a length gauge.\textrm{\ }As is well known,%
\cite{gauge,Sipe,Mer1} in this gauge it is straightforward to study quantum
transitions via intermediate states and to obtain gauge-independent
transition probabilities.

The derived equations are solved numerically for a graphene in the Dirac
cone approximation. Then we consider high harmonic generation process for
moderately strong pump waves and show that one can achieve considerable
enhancement of the harmonic generation rate due to the many-body Coulomb
interaction between the charge carriers.

The paper is organized as follows. In Sec. II the Hamiltonian with many-body
Coulomb interaction in the scope of mean-field approximation and the set of
equations for a single-particle density matrix are formulated. In Sec. III,
we consider multiphoton excitation of Fermi-Dirac sea and generation of
harmonics in graphene. Finally, conclusions are given in Sec. IV.

\section{Evolutionary equation for single-particle density matrix}

Let a graphene monolayer interact with plane quasimonochromatic
electromagnetic wave field. To exclude the effect of wave's magnetic field
we assume that the wave propagates in a perpendicular direction to the
graphene plane ($XY$). Thus, this travelling wave for graphene electrons
becomes a homogeneous quasiperiodic electric field of carrier frequency $%
\omega $ and slowly varying envelope $E_{0}\left( t\right) $. We assume
linearly polarized (along the $x$-axis) wave: 
\begin{equation}
\mathbf{E}\left( t\right) =\widehat{\mathbf{x}}E_{0}\left( t\right) \cos
\omega t.  \label{E_field}
\end{equation}%
The wave amplitude is described by the sin-squared envelope function $%
E_{0}\left( t\right) =E_{0}f\left( t\right) $:%
\begin{equation}
f\left( t\right) =\left\{ 
\begin{array}{cc}
\sin ^{2}\left( \pi t/\mathcal{T}_{p}\right) , & 0\leq t\leq \mathcal{T}_{p},
\\ 
0, & t<0,t>\mathcal{T}_{p},%
\end{array}%
\right.  \label{env}
\end{equation}%
where $\mathcal{T}_{p}$ characterizes the pulse duration. Note that, the
Gaussian and sin-squared envelopes lead to very similar results. The latter
is more convenient for numerical and analytical calculations.\cite{Mil}

Low-energy excitations which are much smaller than the nearest neighbor
hopping energy can be described by an effective Hamiltonian 
\begin{equation}
H_{0}=\hbar \mathrm{v}_{F}\left( 
\begin{array}{cc}
0 & \widehat{k}_{x}-i\widehat{k}_{y} \\ 
\widehat{k}_{x}+i\widehat{k}_{y} & 0%
\end{array}%
\right) ,  \label{DH}
\end{equation}%
where $\mathrm{v}_{F}\approx c/300$ is the Fermi velocity ($c$ is the light
speed in vacuum),\textbf{\ }$\hbar \widehat{\mathbf{k}}$\textbf{\ }is the
electron momentum operator. The eigenstates of the effective Hamiltonian (%
\ref{DH}) are the spinors,%
\begin{equation}
\psi _{\mathbf{k,}\lambda }(\mathbf{r})=\frac{1}{\sqrt{2\mathcal{A}}}\left( 
\begin{array}{c}
e^{-i\theta \left( \mathbf{k}\right) } \\ 
\lambda%
\end{array}%
\right) e^{i\mathbf{kr}},  \label{free}
\end{equation}%
corresponding to energies%
\begin{equation*}
\mathcal{E}_{\lambda }\left( \mathbf{k}\right) =\lambda \hbar \mathrm{v}%
_{F}k.
\end{equation*}%
Here the band index $\lambda =\pm 1$, and $\mathcal{A}$ is the quantization
area, 
\begin{equation}
\theta \left( \mathbf{k}\right) =\arctan \left( \frac{k_{y}}{k_{x}}\right)
\label{angle}
\end{equation}%
is the polar angle in the momentum space. We will work in the second
quantization formalism, expanding the fermionic field operators on the basis
of states given in (\ref{free}), that is,%
\begin{equation}
\widehat{\Psi }(\mathbf{r})=\sum\limits_{\mathbf{k,}\lambda }\widehat{e}%
_{\lambda ,\mathbf{k}}\psi _{\mathbf{k,}\lambda }(\mathbf{r}),  \label{exp}
\end{equation}%
where $\widehat{e}_{\lambda ,\mathbf{k}}$ ($\widehat{e}_{\lambda ,\mathbf{k}%
}^{\dagger }$) is the annihilation (creation) operator for an electron with
momentum $\mathbf{k}$ and band $\lambda $ (for conduction ($\lambda =1$) and
valence ($\lambda =-1$) bands). In (\ref{exp}) we have omitted real spin and
valley quantum numbers because of degeneracy.

The electrons interact through long-range Coulomb forces and the Hamiltonian
for electron-electron interactions can be written in terms of the field
operators, $\widehat{\Psi }(\mathbf{r})$, as:%
\begin{equation*}
\widehat{H}_{\mathrm{c}}=\frac{1}{2}\int d\mathbf{r}\int d\mathbf{r}^{\prime
}\ \widehat{\Psi }^{\dagger }(\mathbf{r})\widehat{\Psi }^{\dagger }(\mathbf{r%
}^{\prime })V_{c}(\mathbf{r}-\mathbf{r}^{\prime })\widehat{\Psi }(\mathbf{r}%
^{\prime })\widehat{\Psi }(\mathbf{r}),
\end{equation*}%
where $V_{c}(\mathbf{r})=e^{2}/\left( \varepsilon \left\vert \mathbf{r}%
\right\vert \right) $ is the bare Coulomb potential, $\varepsilon $ is the
effective dielectric constant of the substrate on which graphene is
deposited.

The light--matter interaction part is taken in the length gauge: 
\begin{equation}
\widehat{H}_{\mathrm{int}}=e\int d\mathbf{r}\ \widehat{\Psi }^{\dagger }(%
\mathbf{r})\mathbf{rE}\left( t\right) \widehat{\Psi }(\mathbf{r}).
\label{LMI}
\end{equation}%
The latter is given in terms of the gauge-independent field $\mathbf{E}%
\left( t\right) $.

Taking into account expansion (\ref{exp}), the total Hamiltonian can be
represented as follow: 
\begin{equation*}
\widehat{H}=\sum_{\lambda ,\mathbf{k}}\mathcal{E}_{\lambda }\left( \mathbf{k}%
\right) \widehat{e}_{\lambda ,\mathbf{k}}^{\dag }\widehat{e}_{\lambda ,%
\mathbf{k}}+H_{\mathrm{Coul}}
\end{equation*}%
\begin{equation*}
+ie\sum_{\lambda ,\mathbf{k,k}^{\prime }}\left( \mathbf{E\cdot }\frac{%
\partial \delta \left( \mathbf{k-k}^{\prime }\right) }{\partial \mathbf{k}%
^{\prime }}\right)
\end{equation*}%
\begin{equation}
\times \left( \mathcal{D}_{\mathbf{k}^{\prime }\mathbf{k}}^{(+)}\widehat{e}%
_{\lambda ,\mathbf{k}^{\prime }}^{\dag }\widehat{e}_{\lambda ,\mathbf{k}}+%
\mathcal{D}_{\mathbf{k}^{\prime }\mathbf{k}}^{(-)}\widehat{e}_{\lambda ,%
\mathbf{k}^{\prime }}^{\dag }\widehat{e}_{-\lambda ,\mathbf{k}}\right) .
\label{Ham12}
\end{equation}%
The Dirac delta function $\delta \left( \mathbf{k-k}^{\prime }\right) $\ in
the light-matter interaction part provides proper inclusion of inter and
intraband transitions.\cite{Sipe,Mer1} The Coulomb interaction reads: 
\begin{equation*}
\widehat{H}_{\mathrm{Coul}}=\frac{1}{2\mathcal{A}}\sum_{_{\lambda
_{1}\lambda _{2}\lambda _{3}\lambda _{4}}}\sum_{\mathbf{q,k,k}^{\prime
}}V_{2D}\left( \mathbf{q}\right) \digamma _{\lambda _{1}\lambda _{2}\lambda
_{3}\lambda _{4}}\left( \mathbf{q,k},\mathbf{k}^{\prime }\right)
\end{equation*}%
\begin{equation}
\times \widehat{e}_{\lambda _{1},\mathbf{k+q}}^{\dag }\widehat{e}_{\lambda
_{2},\mathbf{k}^{\prime }-\mathbf{q}}^{\dag }\widehat{e}_{\lambda _{3},%
\mathbf{k}^{\prime }}\widehat{e}_{\lambda _{4},\mathbf{k}},  \label{CB}
\end{equation}%
where%
\begin{equation}
V_{2D}\left( \mathbf{q}\right) =\frac{2\pi e^{2}}{\varepsilon \left\vert 
\mathbf{q}\right\vert }F\left( \mathbf{q}\right)  \label{2DC}
\end{equation}%
is the 2D Coulomb potential in momentum space and 
\begin{equation*}
\digamma _{\lambda _{1}\lambda _{2}\lambda _{3}\lambda _{4}}\left( \mathbf{%
q,k},\mathbf{k}^{\prime }\right) =\frac{1}{4}\left[ \lambda _{1}\lambda
_{2}\lambda _{3}\lambda _{4}\right.
\end{equation*}%
\begin{equation*}
+e^{i\left[ \theta (\mathbf{k+q})+\theta (\mathbf{k}^{\prime }-\mathbf{q}%
)-\theta (\mathbf{k})-\theta (\mathbf{k}^{\prime })\right] }
\end{equation*}%
\begin{equation}
+\left. \lambda _{2}\lambda _{3}e^{i\left[ \theta (\mathbf{k+q})-\theta (%
\mathbf{k})\right] }+\lambda _{1}\lambda _{4}e^{i\left[ \theta (\mathbf{k}%
^{\prime }-\mathbf{q})-\theta (\mathbf{k}^{\prime })\right] }\right] .
\label{form}
\end{equation}%
In the interaction part of the Hamiltonian (\ref{Ham12}) 
\begin{equation}
\mathcal{D}_{\mathbf{k}^{\prime }\mathbf{k}}^{(\pm )}=\frac{1}{2}\left(
e^{i\left( \theta (\mathbf{k}^{\prime })-\theta (\mathbf{k})\right) }\pm
1\right) .  \label{dm}
\end{equation}%
At that, the term proportional to $\mathcal{D}_{\mathbf{k}^{\prime }\mathbf{k%
}}^{(+)}\mathcal{\ }$is responsible for intraband transitions, while the
term proportional to $\mathcal{D}_{\mathbf{k}^{\prime }\mathbf{k}}^{(-)}$
describes interband transitions.

The Coulomb interaction part (\ref{CB}) contains products of four fermionic
operators. We will treat Coulomb interaction in the scope of mean-field
theory\cite{MFT} reducing the Hamiltonian (\ref{Ham12}) into so-called
mean-field Hamiltonian which allows obtaining closed set of equations for
the dynamic quantities. We need to choose the proper mean field parameters.
Due to the homogeneity of the applied wave-field and initial system, as a
mean field parameters are taken distribution functions for conduction $%
\mathcal{N}_{c}\left( \mathbf{k},t\right) =\left\langle \widehat{e}_{c%
\mathbf{k}}^{\dag }\widehat{e}_{c\mathbf{k}}\right\rangle $ and for valence $%
\mathcal{N}_{v}\left( \mathbf{k},t\right) =\left\langle \widehat{e}_{v%
\mathbf{k}}^{\dag }\widehat{e}_{v\mathbf{k}}\right\rangle $ bands carriers,
and interband polarization $\mathcal{P}\left( \mathbf{k},t\right)
=\left\langle \widehat{e}_{c\mathbf{k}}^{\dag }\widehat{e}_{v\mathbf{k}%
}\right\rangle $. For the Coulomb interaction, this is a Hartree-Fock
approximation. For the mean-field Hamiltonian we will use the following
decompositions:%
\begin{equation*}
\left. \widehat{a}_{\alpha }^{\dag }\widehat{b}_{\beta }^{\dag }\widehat{c}%
_{\gamma }\widehat{d}_{\delta }\right\vert _{\mathrm{Hartree}}=\widehat{a}%
_{\alpha }^{\dag }\widehat{d}_{\delta }\left\langle \widehat{b}_{\beta
}^{\dag }\widehat{c}_{\gamma }\right\rangle +\widehat{b}_{\beta }^{\dag }%
\widehat{c}_{\gamma }\left\langle \widehat{a}_{\alpha }^{\dag }\widehat{d}%
_{\delta }\right\rangle
\end{equation*}%
\begin{equation}
-\left\langle \widehat{a}_{\alpha }^{\dag }\widehat{d}_{\delta
}\right\rangle \left\langle \widehat{b}_{\beta }^{\dag }\widehat{c}_{\gamma
}\right\rangle ,  \label{Har}
\end{equation}%
\begin{equation*}
\left. \widehat{a}_{\alpha }^{\dag }\widehat{b}_{\beta }^{\dag }\widehat{c}%
_{\gamma }\widehat{d}_{\delta }\right\vert _{\mathrm{Fock}}=-\widehat{a}%
_{\alpha }^{\dag }\widehat{c}_{\gamma }\left\langle \widehat{b}_{\beta
}^{\dag }\widehat{d}_{\delta }\right\rangle -\widehat{b}_{\beta }^{\dag }%
\widehat{d}_{\delta }\left\langle \widehat{a}_{\alpha }^{\dag }\widehat{c}%
_{\gamma }\right\rangle
\end{equation*}%
\begin{equation}
+\left\langle \widehat{a}_{\alpha }^{\dag }\widehat{c}_{\gamma
}\right\rangle \left\langle \widehat{b}_{\beta }^{\dag }\widehat{d}_{\delta
}\right\rangle ,  \label{Fock}
\end{equation}%
with the condition 
\begin{equation}
\left\langle \widehat{f}_{\nu }^{\dag }\widehat{g}_{\mu }\right\rangle
=\left\langle \widehat{f}_{\nu }^{\dag }\widehat{g}_{\nu }\right\rangle
\delta _{\nu _{\mu }};\ f,g=a,b,c,d.  \label{cons}
\end{equation}

Taking into account mean-field parameters, the second quantized Hamiltonian (%
\ref{CB}), Eqs. (\ref{Har}), and (\ref{Fock}) for the Coulomb part we have 
\begin{equation*}
\widehat{H}_{\mathrm{MFC}}=-\frac{1}{\mathcal{A}}\sum_{_{\lambda _{1}\lambda
_{2}\lambda _{3}\lambda _{4}}}\sum_{\mathbf{k}^{\prime }\neq \mathbf{k}%
}V_{2D}\left( \mathbf{k}^{\prime }-\mathbf{k}\right) \widetilde{\digamma }%
_{\lambda _{1}\lambda _{2}\lambda _{3}\lambda _{4}}\left( \mathbf{k},\mathbf{%
k}^{\prime }\right)
\end{equation*}%
\begin{equation*}
\times \left\langle \widehat{e}_{\lambda _{3},\mathbf{k}^{\prime }}^{\dag }%
\widehat{e}_{\lambda _{4},\mathbf{k}^{\prime }}\right\rangle \widehat{e}%
_{\lambda _{1},\mathbf{k}}^{\dag }\widehat{e}_{\lambda _{2},\mathbf{k}}
\end{equation*}%
\begin{equation*}
+\frac{1}{2\mathcal{A}}\sum_{_{\lambda _{1}\lambda _{2}\lambda _{3}\lambda
_{4}}}\sum_{\mathbf{k}^{\prime }\neq \mathbf{k}}V_{2D}\left( \mathbf{k}%
^{\prime }-\mathbf{k}\right) \digamma _{\lambda _{1}\lambda _{2}\lambda
_{3}\lambda _{4}}\left( \mathbf{k},\mathbf{k}^{\prime }\right)
\end{equation*}%
\begin{equation}
\times \,\left\langle \,\widehat{e}_{\lambda _{1},\mathbf{k\prime }}^{\dag }%
\widehat{e}_{\lambda _{3},\mathbf{k}^{\prime }}\right\rangle \left\langle 
\widehat{e}_{\lambda _{2},\mathbf{k}}^{\dag }\widehat{e}_{\lambda _{4},%
\mathbf{k}}\right\rangle  \label{mfc}
\end{equation}%
where%
\begin{equation*}
\widetilde{\digamma }_{\lambda _{1}\lambda _{2}\lambda _{3}\lambda
_{4}}\left( \mathbf{k},\mathbf{k}^{\prime }\right) =\digamma _{\lambda
_{1}\lambda _{2}\lambda _{3}\lambda _{4}}\left( \mathbf{k}^{\prime }-\mathbf{%
k,k}^{\prime },\mathbf{k}\right) .
\end{equation*}%
Note that the Hartree contribution $\sim V_{2D}\left( \mathbf{q=0}\right) $
is zero, which is physically related to the charge neutrality of the total
system. The second term in Eq. (\ref{mfc}) is c-number and does not have a
contribution in the equations of motions for $\mathcal{N}_{c}$, $\mathcal{N}%
_{v}$ and $\mathcal{P}$. Rearranging the terms in Eq. (\ref{mfc}) and taking
into account properties of the function (\ref{form}), the mean-field
Hamiltonian for Coulomb interaction can be expressed in the following form\ 
\begin{equation}
\widehat{H}_{\mathrm{MFC}}=\sum_{\lambda \mathbf{k}}\mathcal{\bar{E}}%
\,_{\lambda }\left( \mathbf{k}\right) \widehat{e}_{\lambda ,\mathbf{k}%
}^{\dag }\widehat{e}_{\lambda ,\mathbf{k}}+\sum_{\lambda \mathbf{k}}\Delta
\,_{\lambda }\left( \mathbf{k}\right) \widehat{e}_{\lambda ,\mathbf{k}%
}^{\dag }\widehat{e}_{-\lambda ,\mathbf{k}}+C\mathrm{,}  \label{MFC}
\end{equation}%
where%
\begin{equation*}
\mathcal{\bar{E}}\,_{\lambda }\left( \mathbf{k}\right) =\frac{\lambda }{%
\mathcal{A}}\sum_{\mathbf{k}^{\prime }\neq \mathbf{k}}V_{2D}\left( \mathbf{k}%
-\mathbf{k}^{\prime }\right) \sin [\theta (\mathbf{k})-\theta (\mathbf{k}%
^{\prime })]\mathcal{P}^{\prime \prime }\left( \mathbf{k}^{\prime }\right)
\end{equation*}%
\begin{equation*}
-\frac{\lambda }{2\mathcal{A}}\sum_{\mathbf{k}^{\prime }\neq \mathbf{k}%
}V_{2D}\left( \mathbf{k}-\mathbf{k}^{\prime }\right) \cos [\theta (\mathbf{k}%
)-\theta (\mathbf{k}^{\prime })]\left( \mathcal{N}_{c}\left( \mathbf{k}%
^{\prime }\right) -\mathcal{N}_{v}\left( \mathbf{k}^{\prime }\right) \right)
\end{equation*}%
\begin{equation}
-\frac{1}{2\mathcal{A}}\sum_{\mathbf{k}^{\prime }\neq \mathbf{k}%
}V_{2D}\left( \mathbf{k}-\mathbf{k}^{\prime }\right) \left( \mathcal{N}%
_{c}\left( \mathbf{k}^{\prime }\right) +\mathcal{N}_{v}\left( \mathbf{k}%
^{\prime }\right) \right)  \label{El}
\end{equation}%
is the self-energy corrections due to the electron-electron interactions,
and 
\begin{equation*}
\Delta \,_{\lambda }\left( \mathbf{k}\right) =-\frac{1}{\mathcal{A}}\sum_{%
\mathbf{k}^{\prime }\neq \mathbf{k}}V_{2D}\left( \mathbf{k}-\mathbf{k}%
^{\prime }\right) \mathcal{P}^{\prime }\left( \mathbf{k}^{\prime }\right)
\end{equation*}%
\begin{equation*}
+i\frac{\lambda }{2\mathcal{A}}\sum_{\mathbf{k}^{\prime }\neq \mathbf{k}%
}V_{2D}\left( \mathbf{k}-\mathbf{k}^{\prime }\right) \sin [\theta (\mathbf{k}%
)-\theta (\mathbf{k}^{\prime })]\left( \mathcal{N}_{c}\left( \mathbf{k}%
^{\prime }\right) -\mathcal{N}_{v}\left( \mathbf{k}^{\prime }\right) \right)
\end{equation*}%
\begin{equation}
+i\frac{\lambda }{\mathcal{A}}\sum_{\mathbf{k}^{\prime }}V_{2D}\left( 
\mathbf{k}-\mathbf{k}^{\prime }\right) \cos [\theta (\mathbf{k})-\theta (%
\mathbf{k}^{\prime })]\mathcal{P}^{\prime \prime }\left( \mathbf{k}^{\prime
}\right) .  \label{D1}
\end{equation}%
is the self-polarization corrections. In Eq. (\ref{MFC}) $C$ is the c-number
part of the mean-field Hamiltonian. In Eqs. $\mathcal{P}^{\prime }\left( 
\mathbf{k}\right) $ and $\mathcal{P}^{\prime \prime }\left( \mathbf{k}%
\right) $ are real and imaginary parts of $\mathcal{P}\left( \mathbf{k}%
\right) $, respectively.

Now, from Heisenberg equation%
\begin{equation}
i\hbar \frac{\partial \widehat{e}_{\eta _{2},\mathbf{k}}^{\dag }\widehat{e}%
_{\eta _{1},\mathbf{k}}}{\partial t}=\left[ \widehat{e}_{\eta _{2},\mathbf{k}%
}^{\dag }\widehat{e}_{\eta _{1},\mathbf{k}},\widehat{H}\right] ,
\label{Heiz}
\end{equation}%
where total Hamiltonian $\widehat{H}$ is taken in the mean-field
approximation one can obtain the following evolution equations for $\mathcal{%
N}_{c}\left( \mathbf{k},t\right) $, $\mathcal{N}_{v}\left( \mathbf{k}%
,t\right) $ and $\mathcal{P}\left( \mathbf{k},t\right) $:%
\begin{equation}
\frac{\partial \mathcal{N}_{c}\left( \mathbf{k},t\right) }{\partial t}-\frac{%
e\mathbf{E}}{\hbar }\frac{\partial \mathcal{N}_{c}\left( \mathbf{k},t\right) 
}{\partial \mathbf{k}}=i\Omega _{\mathrm{R-C}}\left( \mathbf{k},t\right) 
\mathcal{P}^{\ast }\left( \mathbf{k},t\right) +\mathrm{c.c.},  \label{ev1}
\end{equation}%
\begin{equation}
\frac{\partial \mathcal{N}_{v}\left( \mathbf{k},t\right) }{\partial t}-\frac{%
e\mathbf{E}}{\hbar }\frac{\partial \mathcal{N}_{v}\left( \mathbf{k},t\right) 
}{\partial \mathbf{k}}=-i\Omega _{\mathrm{R-C}}\left( \mathbf{k},t\right) 
\mathcal{P}^{\ast }\left( \mathbf{k},t\right) +\mathrm{c.c.},  \label{ev2}
\end{equation}%
\begin{equation*}
\frac{\partial \mathcal{P}\left( \mathbf{k},t\right) }{\partial t}-\frac{e%
\mathbf{E}}{\hbar }\frac{\partial \mathcal{P}\left( \mathbf{k},t\right) }{%
\partial \mathbf{k}}=i\omega _{\mathrm{D-C}}\left( \mathbf{k},t\right) 
\mathcal{P}\left( \mathbf{k},t\right)
\end{equation*}%
\begin{equation}
-i\Omega _{\mathrm{R-C}}\left( \mathbf{k},t\right) \left( \mathcal{N}%
_{c}\left( \mathbf{k},t\right) -\mathcal{N}_{v}\left( \mathbf{k},t\right)
\right) ,  \label{ev3}
\end{equation}%
where%
\begin{equation*}
\Omega _{\mathrm{R-C}}\left( \mathbf{k},t\right) =\frac{e\mathbf{E}}{2\hbar }%
\frac{\partial \theta (\mathbf{k})}{\partial \mathbf{k}}-i\frac{1}{2\hbar 
\mathcal{A}}\sum_{\mathbf{k}^{\prime }\neq \mathbf{k}}V_{2D}\left( \mathbf{k}%
-\mathbf{k}^{\prime }\right)
\end{equation*}%
\begin{equation*}
\times \sin [\theta (\mathbf{k})-\theta (\mathbf{k}^{\prime })]\left(
N_{c}\left( \mathbf{k}^{\prime },t\right) -N_{v}\left( \mathbf{k}^{\prime
},t\right) \right)
\end{equation*}%
\begin{equation}
-\frac{1}{\hbar \mathcal{A}}\sum_{\mathbf{k}^{\prime }\neq \mathbf{k}%
}V_{2D}\left( \mathbf{k}-\mathbf{k}^{\prime }\right) \left[ \mathcal{P}%
^{\prime }\left( \mathbf{k}^{\prime },t\right) +i\cos [\theta (\mathbf{k}%
)-\theta (\mathbf{k}^{\prime })]\mathcal{P}^{\prime \prime }\left( \mathbf{k}%
^{\prime },t\right) \right]  \label{Rabi}
\end{equation}%
is the Coulomb corrected Rabi frequency. The Coulomb corrected transition
frequency is defined as%
\begin{equation*}
\omega _{\mathrm{D-C}}\left( \mathbf{k},t\right) =2\mathrm{v}_{F}k+\frac{1}{%
\hbar \mathcal{A}}\sum_{\mathbf{k}^{\prime }\neq \mathbf{k}}V_{2D}\left( 
\mathbf{k}-\mathbf{k}^{\prime }\right)
\end{equation*}%
\begin{equation*}
\times \cos [\theta (\mathbf{k})-\theta (\mathbf{k}^{\prime })]\left( 
\mathcal{N}_{v}\left( \mathbf{k}^{\prime },t\right) -\mathcal{N}_{c}\left( 
\mathbf{k}^{\prime },t\right) \right)
\end{equation*}%
\begin{equation}
+\frac{2}{\hbar \mathcal{A}}\sum_{\mathbf{k}^{\prime }\neq \mathbf{k}%
}V_{2D}\left( \mathbf{k}-\mathbf{k}^{\prime }\right) \sin [\theta (\mathbf{k}%
)-\theta (\mathbf{k}^{\prime })]\mathcal{P}^{\prime \prime }\left( \mathbf{k}%
^{\prime },t\right) .  \label{NPtrF}
\end{equation}

As is seen from Eqs. (\ref{ev1})-(\ref{NPtrF}) in the scope of mean-field
approximation the Coulomb interaction leads to a renormalization of the
light-matter coupling, which depends on $P$ and $N_{c,v}$. Also, the
transition energies become renormalized due to the Coulomb interaction. Note
that, in general, one should also include in Eqs. (\ref{ev1})-(\ref{NPtrF})
the relaxation terms because of carrier-carrier and the carrier-phonon
scatterings\cite{K3}.\textrm{\ }Hence, our consideration is valid in the
ultrafast excitation regime and it is correct only for the times $t<\tau
_{\min }$, where $\tau _{\min }$\ is the minimum of all relaxation times.\
Experiments\cite{K1,K2} and theory\cite{K3} suggest that carrier-carrier
scattering and the carrier-phonon coupling\cite{rel0,rel1,rel2} are the main
relaxation channels for the radiation-excited current in graphene. At the
excitation by 800nm laser, the carrier-carrier scattering results in a
decrease of the current with a decay constant of 100 fs,\cite{K2} that is
almost 40 wave periods.\textrm{\ }Electron-electron interactions in graphene
give rise to a linear energy dependence of the inverse lifetime \cite%
{LT1,LT2}.\textrm{\ }Thus, one can extrapolate this scale to low energy
excitations. The electron-phonon coupling in graphene is considerable for
optical phonons.\textrm{\ }Therefore here we consider excitation with THz
waves far below the energy threshold for the emission of optical phonons ($%
0.2$\ $\mathrm{eV}$\ being a characteristic optical phonon frequency).%
\textrm{\ }In considered case, the electron linewidth due to electron-phonon
interaction is negligible\cite{rel0}, while it increases linearly beyond
this threshold making relaxation times about $1\ \mathrm{ps}$.\cite%
{rel1,rel2} Thus, the pulse duration $\mathcal{T}_{p}$\ is chosen to be $%
\mathcal{T}_{p}=36T$\ , where $T$\ is the wave period, which allows us do
not include relaxation processes.

The obtained equations are closed set of nonlinear integro-differential
equations which should be solved with the proper initial conditions. In the
scope of the mean-field theory one can define the ground state
self-consistently.\cite{MFT,koch,E4}\textrm{\ }In this case, at vanishing
temperatures the ground state is an excitonic condensate with a certain
energy gap \cite{koch,E4}. The gap size is very sensitive to the 2D model of
Coulomb potential and is very small when one takes into account Fermi
velocity renormalization\cite{E4}.\textrm{\ }According to the experimental
results\cite{FV1} the gap is smaller than $\sim 0.1\,\mathrm{meV}$. Note
that tight-binding ground state is also solution of the stationary
self-consistent mean-field equations.\cite{koch}\textrm{\ }Therefore, for
initial state one can assume Fermi-Dirac distribution with the temperature
larger than the predicted excitonic gap: 
\begin{equation}
\mathcal{N}_{v,c}\left( \mathbf{k},0\right) \simeq \frac{1}{1+e^{\frac{\mp
k-k_{F}}{T^{\ast }}}},\ \ \mathcal{P}\left( \mathbf{k},0\right) =0.
\label{equ}
\end{equation}%
Here $k_{F}$ is the Fermi wave number and it is assumed linear dispersion.
The Fermi velocity renormalization is incorporated into the definition of
scaled temperature $T^{\ast }$. For the latter it is assumed $T^{\ast
}=0.1\hbar \omega /\mathrm{v}_{F}.$

For the initial functions (\ref{equ}) $\Omega _{\mathrm{R-C}}\left( \mathbf{k%
},0\right) =0$ and 
\begin{equation*}
\omega _{\mathrm{D-C}}\left( \mathbf{k},0\right) =2\mathrm{v}_{F}k+\frac{1}{%
\hbar \mathcal{A}}\sum_{\mathbf{k}^{\prime }\neq \mathbf{k}}V_{2D}\left( 
\mathbf{k}-\mathbf{k}^{\prime }\right)
\end{equation*}%
\begin{equation}
\times \cos [\theta (\mathbf{k})-\theta (\mathbf{k}^{\prime })]\left( 
\mathcal{N}_{v}\left( \mathbf{k}^{\prime },0\right) -\mathcal{N}_{c}\left( 
\mathbf{k}^{\prime },0\right) \right) .  \label{Self}
\end{equation}%
The latter can be written as%
\begin{equation}
\omega _{\mathrm{D-C}}\left( k,0\right) =2\widetilde{\mathrm{v}}_{F}k,
\label{div0}
\end{equation}%
where%
\begin{equation*}
\widetilde{\mathrm{v}}_{F}=\mathrm{v}_{F}+\frac{e^{2}}{2\pi \varepsilon
\hbar k}
\end{equation*}%
\begin{equation}
\times \int_{0}^{k_{c}}k^{\prime }dk^{\prime }\int_{0}^{2\pi }\cos \theta
d\theta \frac{\left( \mathcal{N}_{v}\left( k^{\prime },0\right) -\mathcal{N}%
_{c}\left( k^{\prime },0\right) \right) }{\sqrt{k^{2}+k^{\prime
2}-2kk^{\prime }\cos \theta }}.  \label{div}
\end{equation}%
is the renormalized Fermi velocity\textrm{. }We note that the integral of
Eq. (\ref{div}) has an ultraviolet high-momentum logarithmic divergence,
which must be regularized through a high wave vector cutoff $k_{c}$, of the
order of the inverse lattice spacing. Conserving the total number of states
in the Brillouin zone, we choose $k_{c}=\left( 4\pi /\mathcal{A}_{c}\right)
^{1/2}$, where $\mathcal{A}_{c}=3\sqrt{3}a^{2}/2$ is the area of the
hexagonal unit cell, and $a=1.42\times 10^{-8}\ \mathrm{cm}$ is the
carbon-carbon distance.

Thus the renormalized frequency can be represented as 
\begin{equation}
\omega _{\mathrm{D-C}}\left( \mathbf{k},t\right) =\omega _{\mathrm{D-C}%
}\left( k,0\right) +\widetilde{\omega }_{\mathrm{D-C}}\left( \mathbf{k}%
,t\right)  \label{wfc}
\end{equation}%
where $\omega _{\mathrm{D-C}}\left( k,0\right) $ is given by the regularized
expression (\ref{div}). Because of finite excitation of Brillouin zone near
the Dirac points now $\widetilde{\omega }_{\mathrm{D-C}}\left( \mathbf{k}%
,t\right) $ and $\Omega _{\mathrm{R-C}}\left( \mathbf{k},t\right) $ are
convergent and one can make integration only near Dirac points.

\section{MULTIPHOTON EXCITATION AND GENERATION OF HARMONICS}

As was mentioned above, equations (\ref{ev1}), (\ref{ev2}), and (\ref{ev3})
are integrodifferential set of nonlinear equations, which can not be solved
analytically. Before numerical solution, one can considerably simplify the
problem. We can make a change of variables and transform the partial
differential equations into ordinary ones. The new variables are $t$ and $%
\widetilde{\mathbf{k}}=\mathbf{k}-\mathbf{k}_{E}$ $\left( t\right) $, where
\ 
\begin{equation*}
\hbar \mathbf{k}_{E}\left( t\right) =-e\int_{0}^{t}\mathbf{E}\left(
t^{\prime }\right) dt^{\prime }
\end{equation*}%
is the classical momentum given by the wave field. After these
transformations, the integration of equations (\ref{ev1})-(\ref{ev3}) is
performed on a grid of $6000-20000$ ($\widetilde{k}_{0},\theta _{0}$)-points
depending on the intensity of the pump wave. For the integration over polar
angle, we use Gaussian quadrature with $60$ points. For $\widetilde{k}_{0}$
we take points homogeneously distributed between the points $\widetilde{k}%
_{0}=0$ and $\widetilde{k}_{0}=\alpha \omega _{0}/\mathrm{v}_{F}$, where $%
\alpha $ depends on the intensity of the pump wave. The time integration is
performed with the standard fourth-order Runge-Kutta algorithm. 
\begin{figure}[tbp]
\includegraphics[width=.5\textwidth]{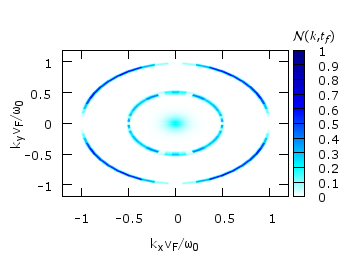}
\caption{(Color online) Electron distribution function $N_{c}\left( \mathbf{k%
},t_{f}\right) $ (in arbitrary units) after the interaction at the instant $%
t_{f}=36\mathcal{T}$, as a function of dimensionless momentum components.
The Coulomb interaction parameter $\protect\alpha _{g}=0$. The wave-particle
dimensionless interaction parameter is taken to be $\protect\chi _{0}=0.3$. }
\end{figure}
\begin{figure}[tbp]
\includegraphics[width=.5\textwidth]{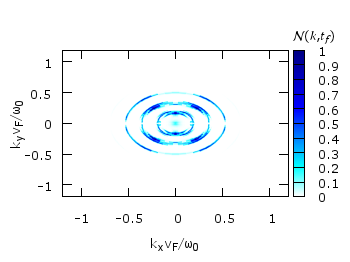}
\caption{(Color online) Same as Fig. 1 but for $\protect\alpha _{g}=2.2$.}
\end{figure}

In graphene, because of the linear scaling of the kinetic energy with
momentum, the ratio of Coulomb to kinetic energy is independent of the
electronic density and equals to $\alpha _{g}=e^{2}/\left( \varepsilon \hbar 
\mathrm{v}_{F}\right) $ depending only on material properties and
environmental conditions. For freestanding graphene ($\varepsilon =1$) $%
\alpha _{g}\approx 2.2$. In most of the experiments, graphene lies on top of
some substrate. In particular, for substrate $\mathrm{SiO}_{2}$ the Coulomb
interaction is moderate. For graphene in contact with air and $\mathrm{SiO}%
_{2}$, $\varepsilon \approx 2.75$ and for interaction parameter we have $%
\alpha _{g}\approx 0.8$. The background dielectric constant can be
significantly enhanced in the presence of substrates in contact with strong
dielectric liquids such as ethanol ($\varepsilon \approx 13$). Thus, in the
experiment one can change $\varepsilon \ $and, as a consequence, to tune
Coulomb interaction. 
\begin{figure}[tbp]
\includegraphics[width=.5\textwidth]{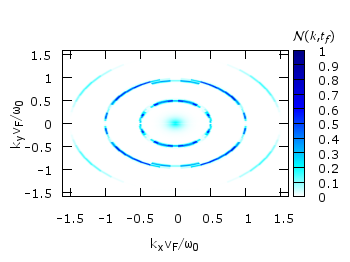}
\caption{(Color online) Electron distribution function $N_{c}\left( \mathbf{k%
},t_{f}\right) $ after the interaction as a function of dimensionless
momentum components. The Coulomb interaction parameter $\protect\alpha %
_{g}=0 $. The wave-particle dimensionless interaction parameter is taken to
be $\protect\chi _{0}=0.5$.}
\end{figure}
\begin{figure}[tbp]
\includegraphics[width=.5\textwidth]{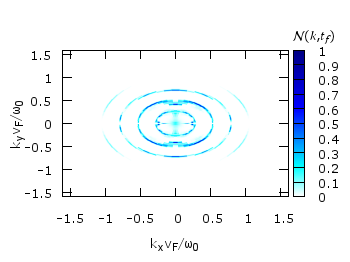}
\caption{(Color online) Same as Fig. 3 but for the Coulomb interaction
parameter $\protect\alpha _{g}=0.8$. }
\end{figure}
In graphene, the wave-particle interaction is characterized by the
dimensionless parameter\cite{Mer1,Mer2}%
\begin{equation*}
\chi _{0}=\frac{eE_{0}\mathrm{v}_{F}}{\omega _{0}}\frac{1}{\hbar \omega _{0}}%
,
\end{equation*}%
which represents the work of the wave electric field $E_{0}$ on a period $%
1/\omega _{0}$ in the units of photon energy $\hbar \omega _{0}$. Here we
consider moderately strong pump waves $\chi _{0}\lesssim 1$.
Photoexcitations of the Fermi-Dirac sea are presented in Figs. 1--4. In Fig.
1 and 2 density plot of the particle distribution function $N_{c}\left( 
\mathbf{k},t_{f}\right) $ after the interaction at the instant $t_{f}=36%
\mathcal{T}$, as a function of dimensionless momentum components are shown.
In both figures the wave-particle dimensionless interaction parameter $\chi
_{0}$ is taken to be $\chi _{0}=0.3$. As is seen from these figures,
depending on the strength of the Coulomb interaction, the photoexcitation
picture changes significantly. Thus, for $\alpha _{g}=0$ the main
contribution is conditioned by the one and two-photon transitions. For $%
\alpha _{g}=2.2$ it is clearly seen the three-photon transition. This is a
consequence of Coulomb interaction on the quasienergy spectrum. Thus, the
multiphoton probabilities of particle-hole pair production will have maximal
values for the resonant transitions%
\begin{equation*}
\overline{\omega }\left( \mathbf{k}_{0}\right) =n\hbar \omega ,\ \
n=1,2,3...,
\end{equation*}%
where%
\begin{equation*}
\overline{\omega }\left( \mathbf{k}_{0}\right) =\frac{1}{\mathcal{T}}%
\int\limits_{0}^{\mathcal{T}}\omega _{\mathrm{D-C}}\left( \mathbf{k}_{0}+%
\mathbf{k}_{E}\left( t\right) ,t\right) dt
\end{equation*}%
is the mean value of the Coulomb and wave-fields dressed transition
frequency (\ref{wfc}). In particular, effective Fermi velocity increases
because of self-energy corrections due to the electron-electron
interactions, and as a consequence resonant wave numbers are decreased.
Besides, due to the enhancement of the effective interaction parameter, we
see in Fig. 2 three-photon transitions. The same picture we see in Figs. 3
and 4 for slightly higher intensities but for moderate Coulomb interaction
parameter $\alpha _{g}=0.8$. Here at $\chi _{0}=0.5$ four-photon transition
is effective due to electron-electron Coulomb interaction.

We further examine the nonlinear response of graphene considering the
generation of harmonics from the multiphoton excited states. At the
multiphoton excitation, particle-hole annihilation will cause intense
coherent radiation of the harmonics of the applied wave field. For the
coherent part of the radiation spectrum, one needs the mean value of the
current density operator. The optical excitation via a linearly polarized
coherent radiation pulse induces transitions in the Fermi-Dirac sea which
results in the surface current in the polarization direction of the pump wave%
\begin{equation*}
\mathcal{J}_{x}\left( t\right) =-\frac{eg_{s}g_{v}}{(2\pi )^{2}}\int d%
\mathbf{k}\widetilde{\mathrm{v}}_{F}\left[ \cos \theta (\mathbf{k})\left( 
\mathcal{N}_{c}\left( \mathbf{k},t\right) -\mathcal{N}_{v}\left( \mathbf{k}%
,t\right) \right) \right.
\end{equation*}%
\begin{equation}
\left. +2\sin \Theta \left( \mathbf{p}\right) \mathcal{P}^{\prime \prime
}\left( \mathbf{k},t\right) \right] ,  \label{curr1}
\end{equation}%
where $g_{s}=2$\ and $g_{v}=2$\ are the spin and valley degeneracy factors,
respectively. This current has a nonlinear dependence on the pump wave
field. At that, due to the graphene symmetries, one can expect intense
radiation of odd harmonics of the incoming wave-field. The harmonics will be
described by the additional generated fields $E_{x}^{(g)}$. We assume that
the generated fields are considerably smaller than the incoming field $%
\left\vert E_{x}^{(g)}\right\vert \ll \left\vert E\right\vert $. In this
case, we can solve Maxwell's wave equation in the propagation direction with
the given source term:%
\begin{equation}
\frac{\partial ^{2}E_{x}^{(t)}}{\partial z^{2}}-\frac{1}{c^{2}}\frac{%
\partial ^{2}E_{x}^{(t)}}{\partial t^{2}}=\frac{4\pi }{c^{2}}\frac{\partial 
\mathcal{J}_{x}\left( t\right) }{\partial t}\delta \left( z\right) .
\label{Max}
\end{equation}%
Here $\delta \left( z\right) $ is the Dirac delta function ($z=0$ is the
graphene plane), $E_{x}^{(t)}$ is the total field. The solution to equation (%
\ref{Max}) reads%
\begin{equation*}
E_{x}^{(t)}\left( t,z\right) =E_{x}\left( t-z/c\right)
\end{equation*}%
\begin{equation}
-\frac{2\pi }{c}\left[ \theta \left( z\right) \mathcal{J}_{x}\left(
t-z/c\right) +\theta \left( -z\right) \mathcal{J}_{x}\left( t+z/c\right) %
\right] ,  \label{sol}
\end{equation}%
where $\theta \left( z\right) $ is the Heaviside step function. The first
term in Eq. (\ref{sol}) is the pump wave. From Eq. (\ref{sol}), we see that
after the encounter with the graphene sheet two propagating waves are
generated. One traveling in the propagation direction of the incoming pulse
and one traveling in the opposite direction. We assume that the spectrum is
measured at a fixed observation point in the forward propagation direction.
For the generated field at $z>0$ we have%
\begin{equation}
E_{x}^{(g)}\left( t-z/c\right) =-\frac{2\pi }{c}\mathcal{J}_{x}\left(
t-z/c\right) .  \label{solut}
\end{equation}%
Thus, solving Eqs. (\ref{ev1}), (\ref{ev2}), and (\ref{ev3}) with the
initial condition (\ref{equ}) and making integration in Eq. (\ref{curr1}),
one can reveal nonlinear response of the graphene.

Before proceeding to the high harmonic generation process we will analyze
the impact of Coulomb interaction on the electromagnetic response of
graphene when the perturbation theory is valid. We will consider linear and
third order response of graphene assuming infinite pulse ($f\left( t\right)
=1$). According to perturbation theory\textrm{\ }%
\begin{equation}
\mathcal{J}_{x}\left( t\right) =\sigma ^{(1)}\frac{E_{0}}{2}e^{i\omega
_{0}t}+\sigma ^{(3)}\left( \frac{E_{0}}{2}\right) ^{3}e^{i3\omega _{0}t}+%
\mathrm{c.c.,}  \label{pert}
\end{equation}%
where $\sigma ^{(1)}$\ and $\sigma ^{(3)}$\ are the linear and third order
conductivities, respectively. In Figure 5 we show the real part of the
linear conductivity. As is seen from Fig. 5, for small $\alpha _{g}$ the
linear conductivity is close to the universal value $\sigma
_{0}=e^{2}/\left( 4\hbar \right) $. However, it increases for the large
interaction parameter $\alpha _{g}$. The latter is consistent with the
perturbative result of the linear response of graphene.\cite{clop2} The
third order conductivity strongly depends on the pump frequency ($\omega
_{0}^{-4}$), so in Fig. 6 we plot $\left\vert \sigma ^{(3)}\right\vert $\
normalized to $\kappa =e^{4}\mathrm{v}_{F}^{2}/(\hbar ^{3}\omega _{0}^{4})$.
As is seen from Fig. 6, the third order conductivity also increases with the
increase of the interaction parameter.\cite{3rd8} As will be seen further,
this tendency holds also for high harmonics. 
\begin{figure}[tbp]
\includegraphics[width=.43\textwidth]{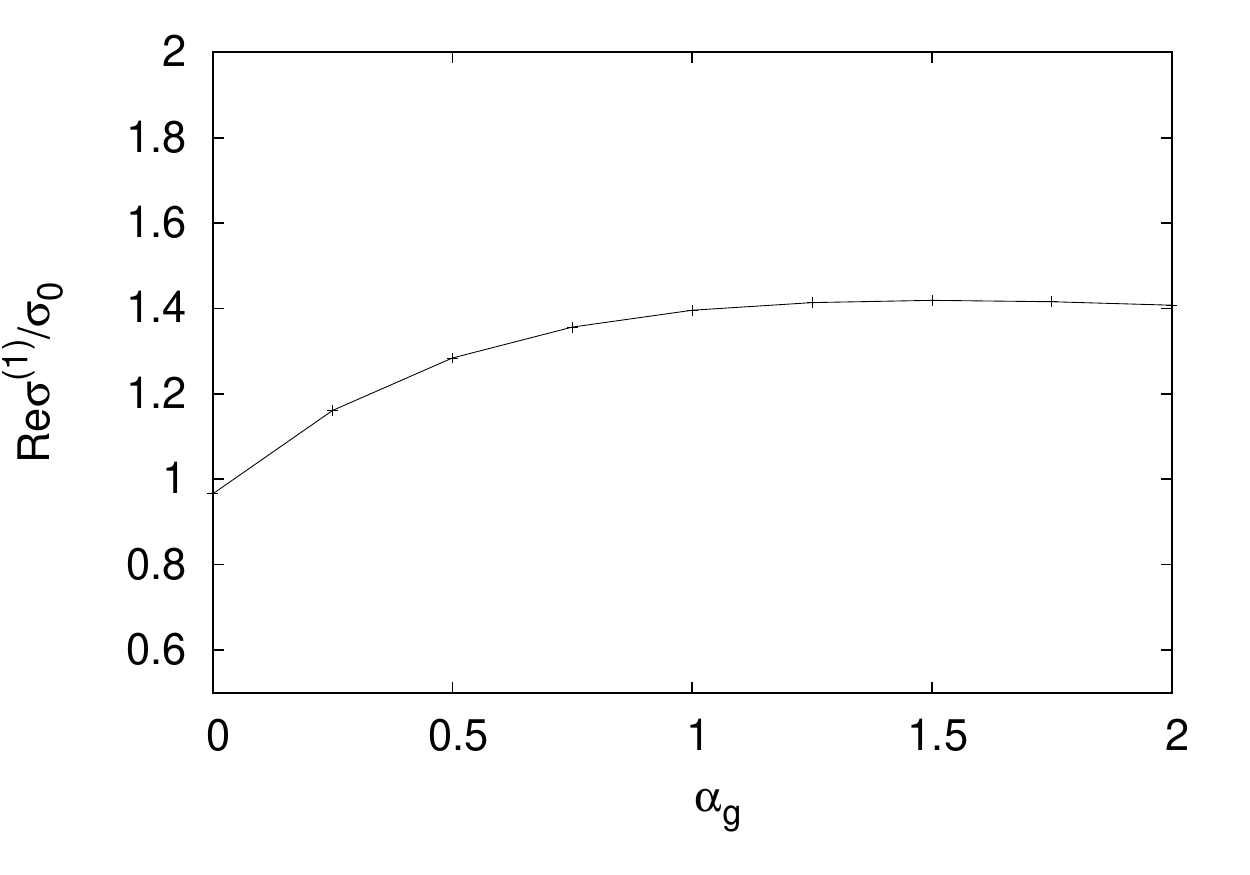}
\caption{Real part of the linear optical conductivity of graphene at $k_{F}=0
$ versus $\protect\alpha _{g}$. The pump wavelength is taken to be $\protect%
\lambda _{0}=0.01\ \mathrm{cm}$. }
\end{figure}
\begin{figure}[tbp]
\includegraphics[width=.43\textwidth]{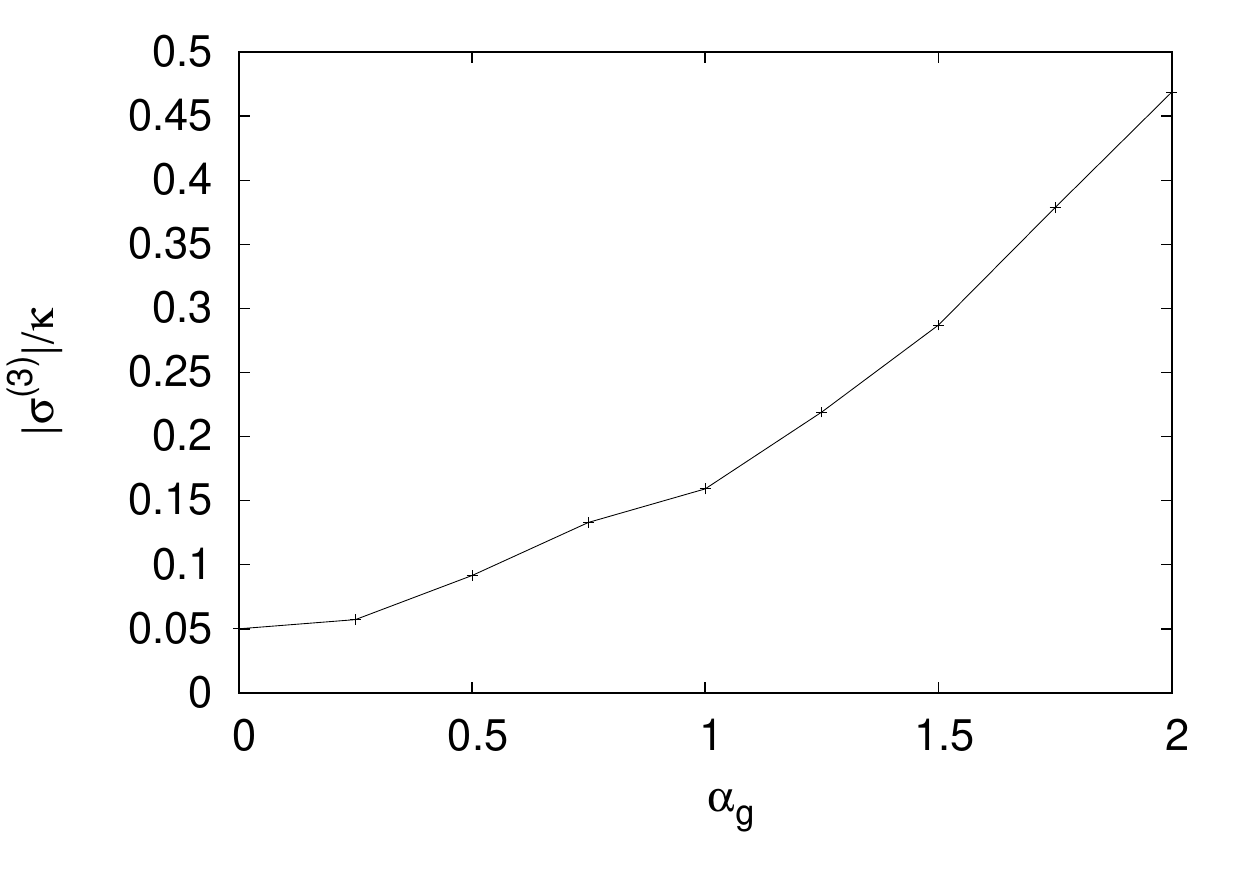}
\caption{The third order conductivity of graphene normalized to $\protect%
\kappa =e^{4}\mathrm{v}_{F}^{2}/(\hbar ^{3}\protect\omega _{0}^{4})$ at $%
k_{F}=0$ versus $\protect\alpha _{g}$. The pump wavelength is taken to be $%
\protect\lambda _{0}=0.01\ \mathrm{cm}$. }
\end{figure}

Making Fourier transform of the generated field (\ref{solut}), one can
calculate the strength of the harmonics. The emission strength of the $s$th
harmonic will be characterized by the dimensionless parameter 
\begin{equation}
\chi _{s}=\frac{e\left\vert E_{x}^{(g)}\left( s\right) \right\vert \mathrm{v}%
_{F}}{\hbar \omega _{0}^{2}}=\chi _{0}\frac{\left\vert E_{x}^{(g)}\left(
s\right) \right\vert }{E_{0}},  \label{khi}
\end{equation}%
where%
\begin{equation}
E_{x}^{(g)}\left( s\right) =\frac{\omega _{0}}{2\pi }\int_{0}^{2\pi /\omega
_{0}}E_{x}^{(g)}\left( t\right) e^{is\omega _{0}t}dt.  \label{F1}
\end{equation}%
With the fast Fourier transform algorithm instead of discrete functions $%
\chi _{s}$ we calculate smooth function $\chi \left( \omega \right) $ and so 
$\chi _{s}=\chi \left( s\omega _{0}\right) $. 
\begin{figure}[tbp]
\includegraphics[width=.5\textwidth]{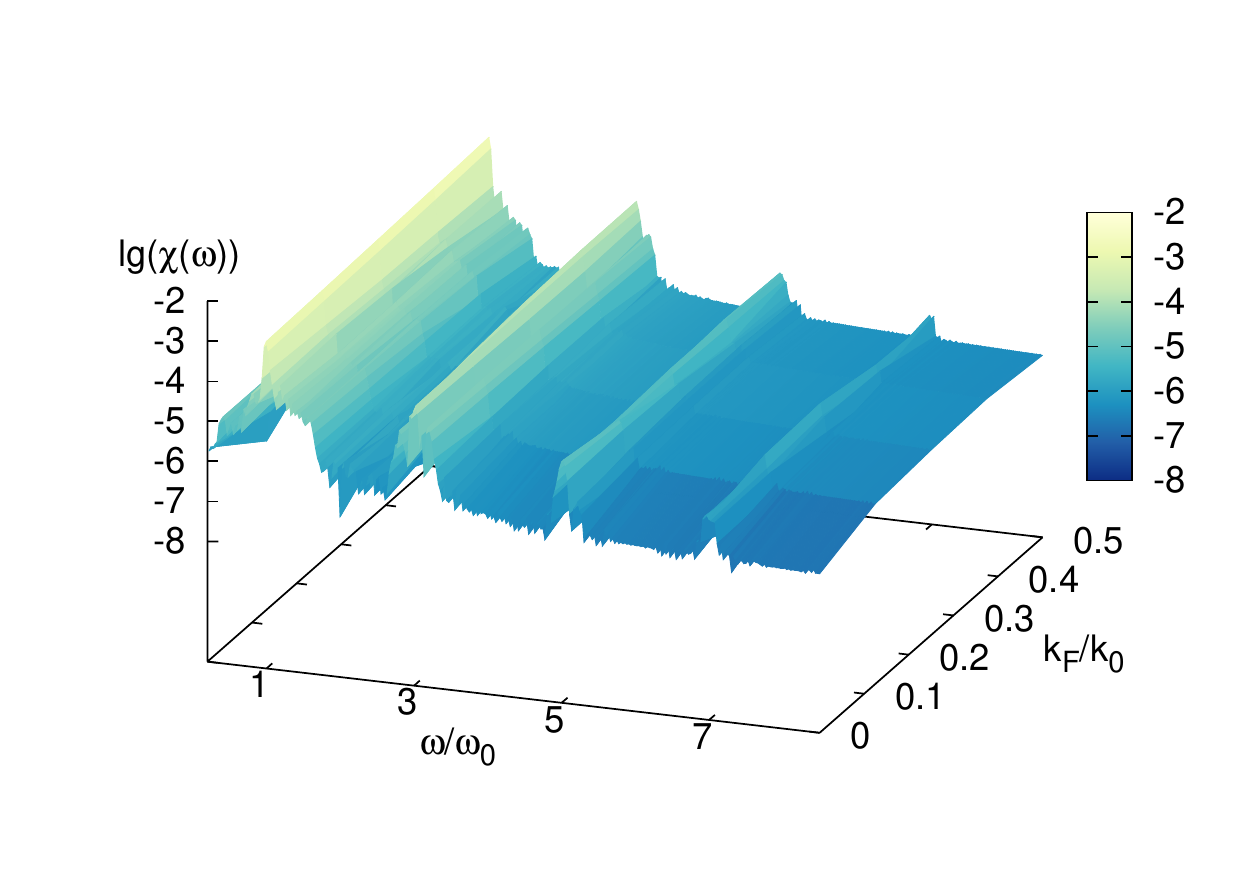}
\caption{(Color online) The radiation spectrum via logarithm of the
normalized field strength $\protect\chi \left( \protect\omega \right) $ (in
arbitrary units) versus Fermi wave number. The pump wavelength is taken to
be $\protect\lambda _{0}=0.01\ \mathrm{cm}$. The wave-particle dimensionless
interaction parameter is taken to be $\protect\chi _{0}=0.3$. The Coulomb
interaction parameter $\protect\alpha _{g}=2.2$.}
\end{figure}
\begin{figure}[tbp]
\includegraphics[width=.5\textwidth]{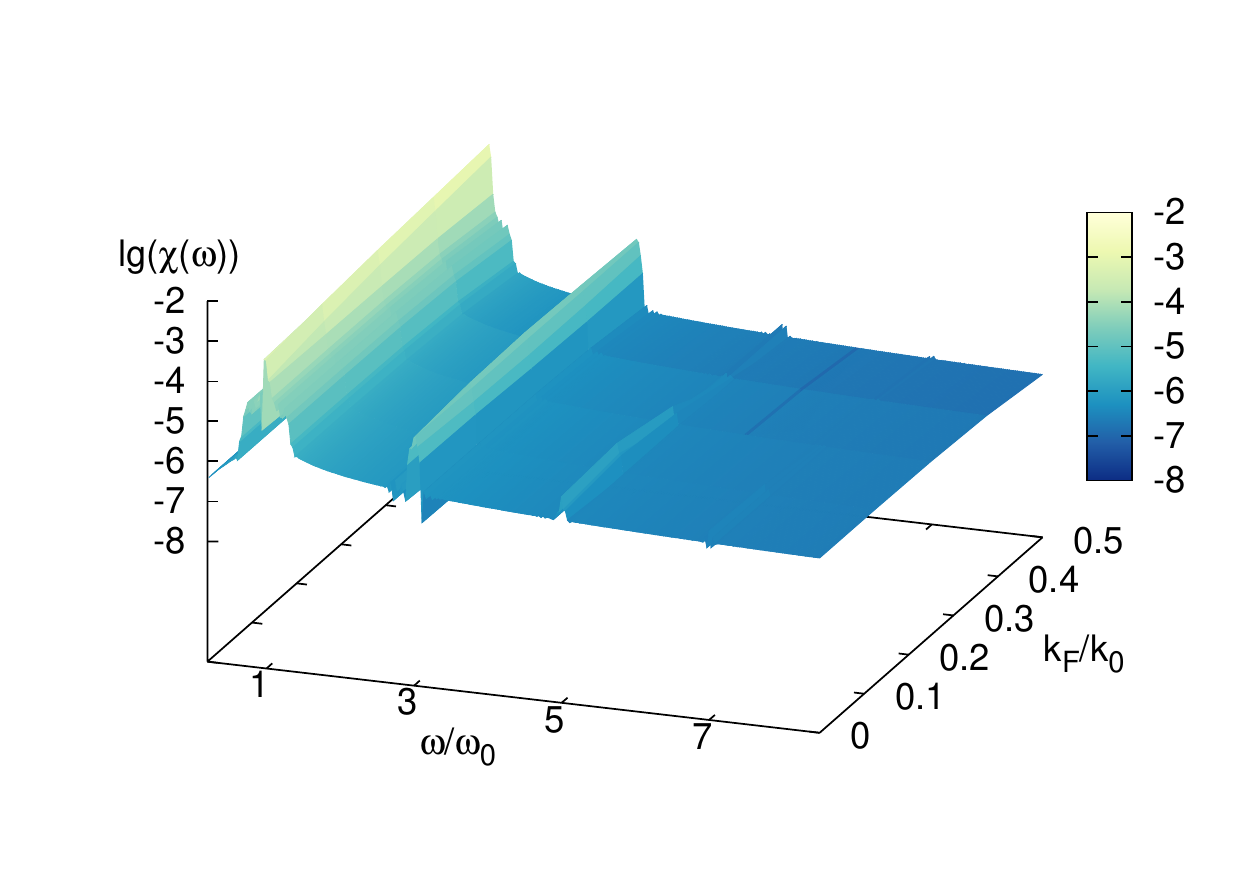}
\caption{(Color online) Same as Fig. 5 but for the Coulomb interaction
parameter $\protect\alpha _{g}=0$.}
\end{figure}
Figures 7 and 8 show the radiation spectrum via logarithm of the normalized
field strength $\chi \left( \omega \right) $ for $\alpha _{g}=2.2$ and $%
\alpha _{g}=0$, respectively. Here we plot $\chi \left( \omega \right) $
versus Fermi wave number. The latter is normalized to $k_{0}=\omega _{0}/%
\mathrm{v}_{F}$. The pump wavelength is taken to be $\lambda _{0}=0.01\ 
\mathrm{cm}$. Comparing Figs. 7 and 8 we see the strong influence of the
Coulomb interaction on the high harmonics radiation spectrum. In Fig. 7 with
strong Coulomb interaction, 5th and 7th harmonics appear, while at $\alpha
_{g}=0$ only 3rd harmonic is feasible. We also see that due to Coulomb
interaction the peaks are broadened.

We also have made calculations for moderate Coulomb interaction parameter $%
\alpha _{g}=0.8$. The results are shown in Figures 9 and 10. In this case,
we also see the enhancement of harmonics order due to Coulomb interaction. 
\begin{figure}[tbp]
\includegraphics[width=.5\textwidth]{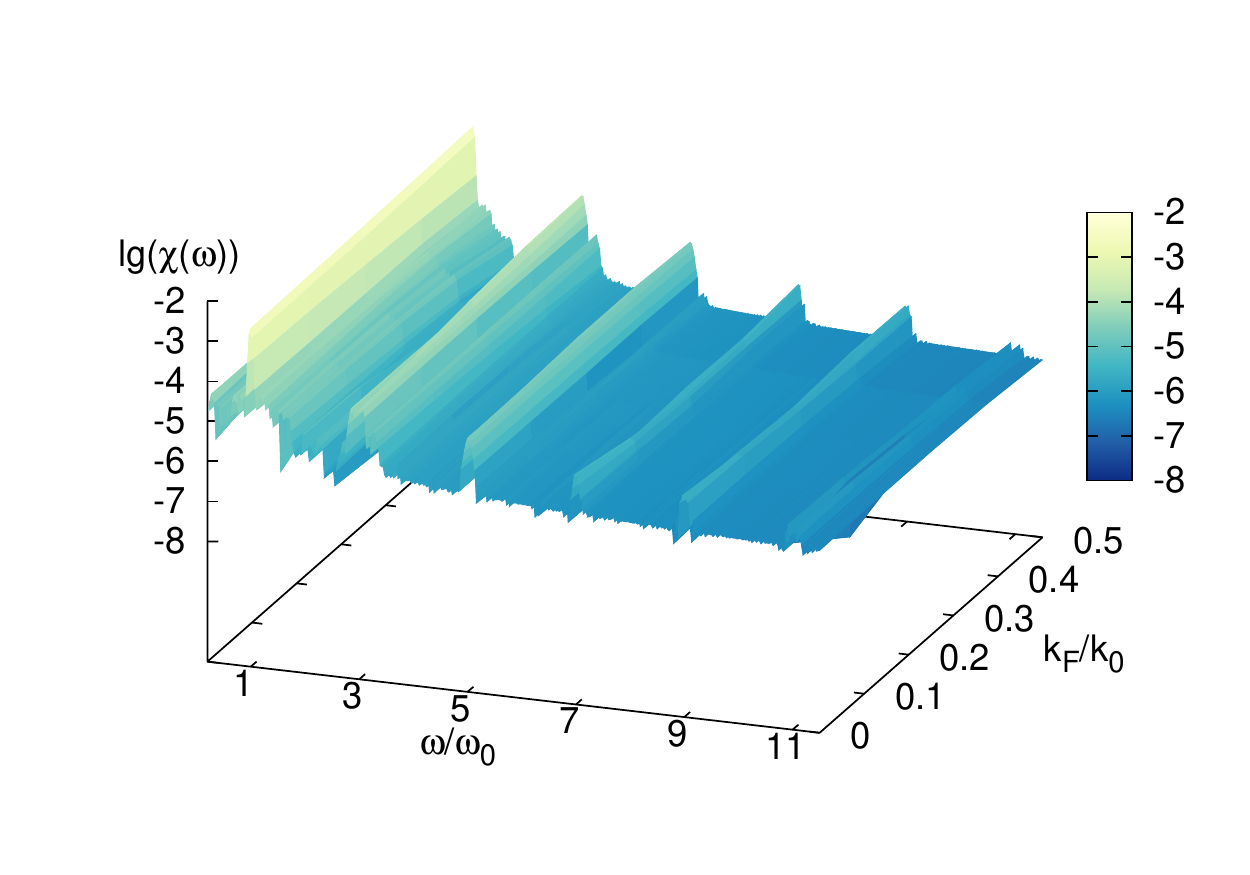}
\caption{(Color online) The radiation spectrum via logarithm of the
normalized field strength $\protect\chi \left( \protect\omega \right) $ (in
arbitrary units) versus Fermi wave number. The pump wavelength is taken to
be $\protect\lambda _{0}=0.1\ \mathrm{cm}$. The wave-particle dimensionless
interaction parameter is taken to be $\protect\chi _{0}=0.5$. The Coulomb
interaction parameter $\protect\alpha _{g}=0.8$.}
\end{figure}
\begin{figure}[tbp]
\includegraphics[width=.5\textwidth]{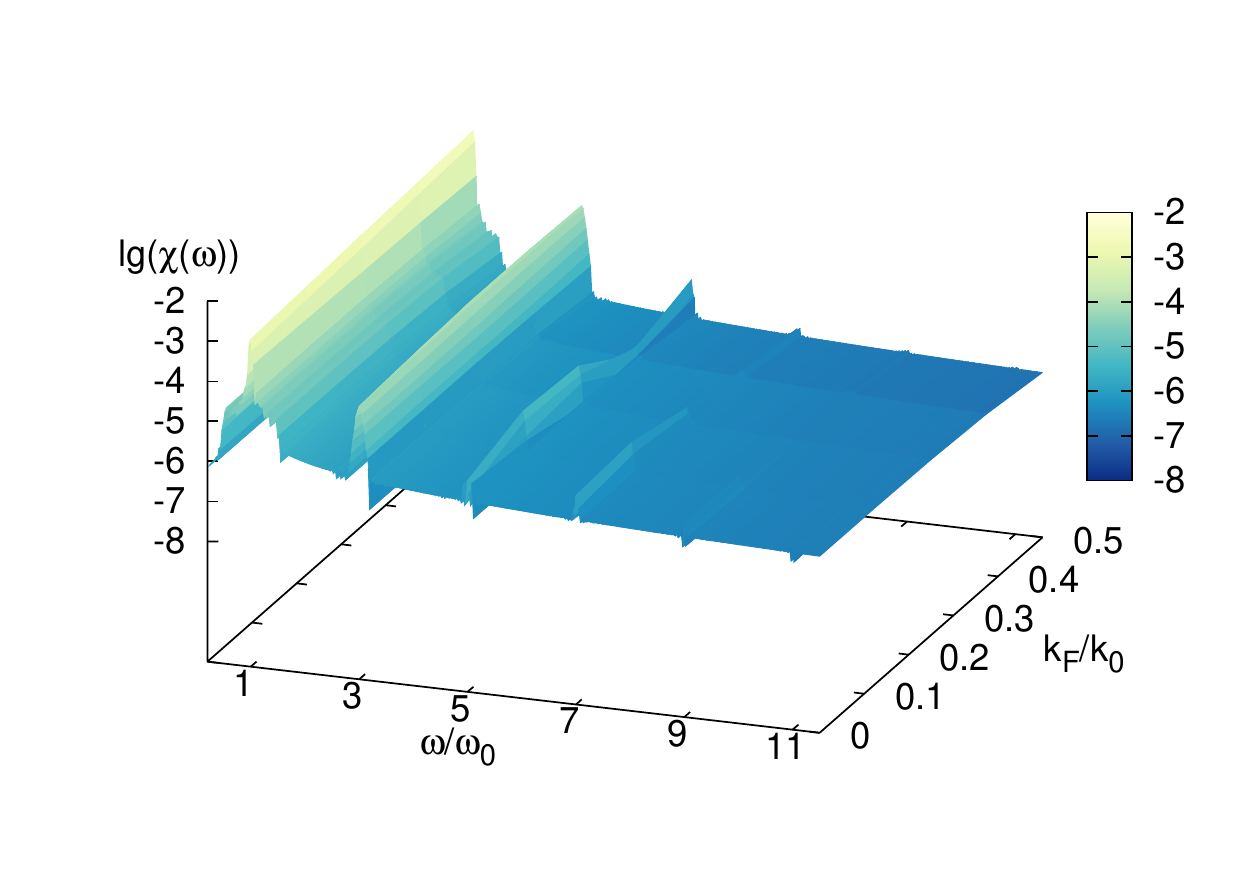}
\caption{(Color online) Same as Fig. 7 but for the Coulomb interaction
parameter $\protect\alpha _{g}=0$.}
\end{figure}
In Fig. 11 we plot high harmonics generation rate depending on the Coulomb
interaction parameter at the\ fixed values of the Fermi wave number ($%
k_{F}=0 $) and pump wave frequency. As is seen from Fig. 11, we have an
increase of the harmonics emission rates at the large Coulomb interaction.

\begin{figure}[tbp]
\includegraphics[width=.5\textwidth]{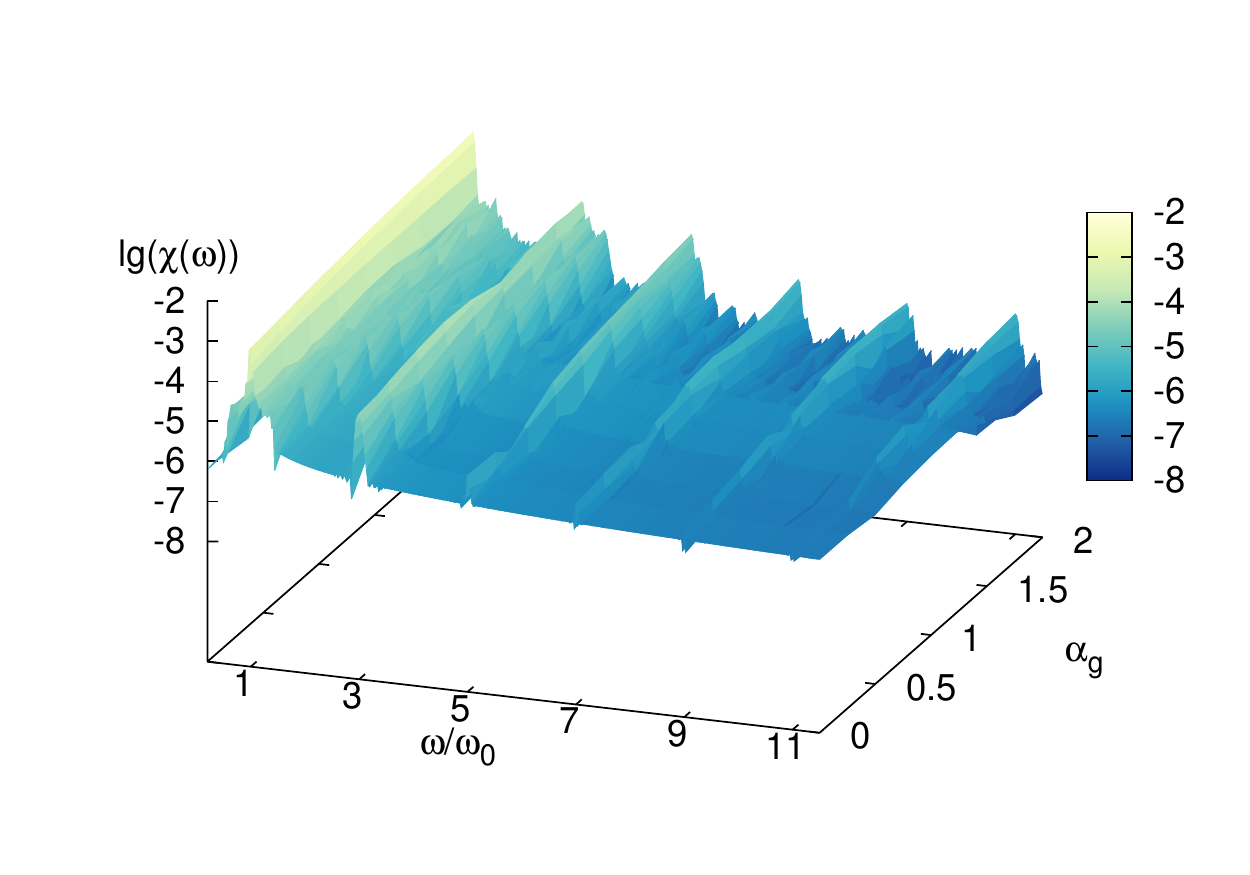}
\caption{(Color online) The radiation spectrum via logarithm of the
normalized field strength $\protect\chi \left( \protect\omega \right) $ (in
arbitrary units) versus interaction parameter $\protect\alpha _{g}$. The
pump wavelength is taken to be $\protect\lambda _{0}=0.1\ \mathrm{cm}$. The
wave-particle dimensionless interaction parameter is taken to be $\protect%
\chi _{0}=0.5$.}
\end{figure}

We also examine how the revealed picture behaves depending on the pump wave
frequency at $k_{F}=0$. The results of our calculations are shown in Fig.
12. Thus, at moderately strong pump waves for the broad range of frequencies
we have intense radiation of harmonics due to Coulomb interaction. 
\begin{figure}[tbp]
\includegraphics[width=.5\textwidth]{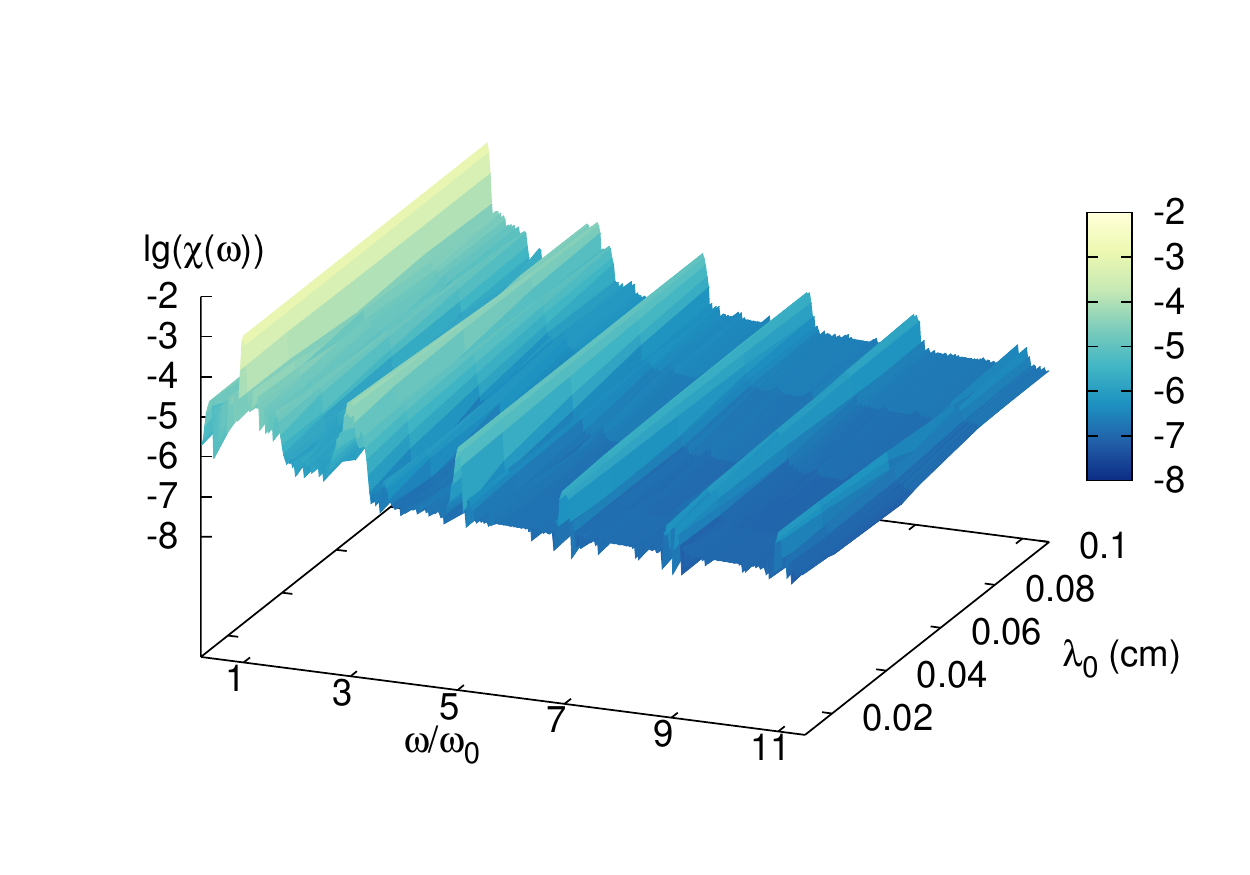}
\caption{(Color online) The radiation spectrum via logarithm of the
normalized field strength $\protect\chi \left( \protect\omega \right) $ (in
arbitrary units) versus pump wavelength. The Fermi wave number is taken to
be $k_{F}=0$. The dimensionless interaction parameter is taken to be $%
\protect\chi _{0}=0.5$. The Coulomb interaction parameter $\protect\alpha %
_{g}=0.8$.}
\end{figure}

For the stronger pump waves $\chi _{0}>1$ one should\ increase integration
domain in Eqs. (\ref{Rabi}), (\ref{NPtrF}), and decrease time step, which
considerably enhance computation time and requires calculations on the
supercomputer. However, taking into account above reported results, one can
definitely state that at the consideration of high-harmonics generation in
graphene one should take into account collective electron-electron
interaction.

We have taken into account the electron-electron interaction in the
Hartree-Fock approximation. The latter is justified if the characteristic
Coulomb interaction energy is smaller than the kinetic energy of electrons.%
\textrm{\ }For the massless particles the ratio of interaction energy to the
kinetic energy, as mentioned above -does not depend on the electronic
density, and is $\alpha _{g}$. However, after the renormalization of Fermi
velocity (\ref{div}) here the effective interaction parameter is $\widetilde{%
\alpha }_{g}=e^{2}/\left( \varepsilon \hbar \widetilde{\mathrm{v}}%
_{F}\right) <1$. Hence, the Hartree-Fock approximation is justified.\textrm{%
\ }Note that we have also considered a case of small Fermi wave numbers%
\textrm{. }Otherwise, one should make Hartree-Fock approximation with the
screened Coulomb potential\cite{K3}.

\section{Conclusion}

We have presented the microscopic theory of nonlinear interaction of the
monolayer graphene with strong coherent radiation field taking into account
many-body electron-electron Coulomb interaction. For the Coulomb
interaction, we have used the self-consistent Hartree-Fock approximation
that leads to a closed set of integrodifferential equations for the
single-particle density matrix. The latter is solved numerically for
graphene in the Dirac cone approximation and ultrafast excitation regime.
For the pump wave, THz frequency range has been taken. We have considered
multiphoton excitation of Fermi-Dirac sea towards the high harmonics
generation. It has been shown that the role of Coulomb interaction in the
nonlinear optical response of graphene is quite considerable that persist
for a wide range of the pump wave frequencies and intensities. Numerical
calculations show that one can reach the efficient generation of high
harmonics with radiation fields of moderate intensities due to Coulomb
mediated enhancement of harmonics order. Because of limited computation
resources, we have made calculations for $\chi _{0}<1$. However, our results
show that at least in the THz range of pump wave frequencies and at the $%
\alpha _{g}>0.5$ one should take into consideration many-body Coulomb
interaction for investigation of the nonlinear optical response of graphene.

\begin{acknowledgments}
This work was supported by the RA MES State Committee of Science and
Belarusian Republican Foundation for Fundamental Research (RB) in the frames
of the joint research projects SCS AB16-19 and BRFFR F17ARM-25, accordingly.
\end{acknowledgments}

\end{document}